# Optimizing Order Dispatch Decisions under Delivery Window Constraints

Khalid Y. Aram

School of Business & Technology, Emporia State University, Emporia, KS 66801, USA

January 2024

**Abstract**

This study focuses on order dispatch decisions within two-echelon supply chains, where order dispatch creates economic shipments to reduce delivery costs. Dispatching orders is often constrained by delivery windows, leading to penalty costs for untimely deliveries. Prolonged dispatch times can increase the lead time of orders and potentially violate these delivery windows. To balance the trade-offs between lead time and economic delivery, this study introduces a simulation-optimization approach for determining optimal ordering and dispatch rules. It emphasizes the intricacies of the order dispatch process and explores how these can be integrated into the simulation-optimization procedure to improve ordering and delivery decisions. The study evaluates various options for implementing dispatch rules, including the number of dispatch queues and prioritized dispatch. The results indicate that a single-queue, quantity-based, first-in-first-out dispatch approach achieves the greatest cost reduction while maintaining a desirable service level.

## 1. Introduction

Lead time of orders includes the time needed to form and dispatch economic-size shipments (Çetinkaya et al., 2006). The dispatch process takes place after fulfilling orders from downstream customers (e.g., retailer). The consolidation of orders reduces transportation cost and emissions resulting from longer delivery trips; however, it increases the risk of extending the lead time of orders. Prolonged lead time of orders has negative impact on customer satisfaction and customer-supplier relationship (Xiang & Rossetti, 2014). Further, delayed delivery of orders may incur penalty cost when the supplier works under a delivery-time constraint, e.g., a delivery time window. Delivery time window of an order is set based on an agreement between the supplier and the customer. The agreement provides that the supplier must deliver the customer order within a specified period. An illustration of a delivery time window is shown in figure 1. *f(x)* in figure 1 is the probability density function of order lead time, C1 is the earliest delivery time allowed, and DC is the width of the delivery window. Delivery windows were initially used in job-shop scheduling to improve timeliness of job completion (Liman & Ramaswamy, 1994). Delivery windows were then applied to improve delivery timeliness in supply chains, increase supplier reliability, and emphasize customer-supplier relationships (Guiffrida & Nagi, 2006).

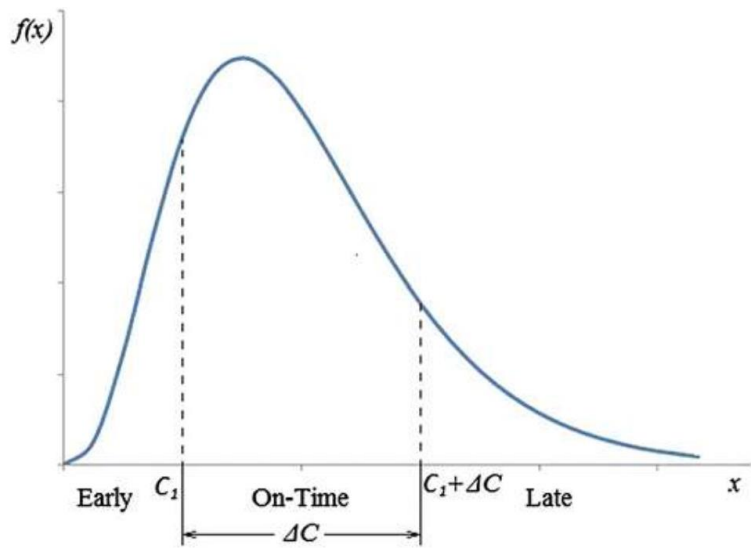

**Figure 1:** Delivery time window of an order (Jaruphongsa et al., 2004).

Under the described situation, the supplier must handle transportation limitations and delivery window constraints at the same time. Creating a balance between the two requirements can help in reducing the total inventory cost and maintaining a desirable customer service level (Bushuev & Guiffrida, 2012). One way to establish this balance can be designed through optimizing the order dispatch process, which is the focus of this study.

Dispatch policies can take different forms. Çetinkaya et al. (2006) described three dispatch policies: quantity-based, schedule-based, and mixed (hybrid). According to a quantity-based policy, a delivery is made when the total size of accumulated orders exceeds a certain threshold. Under a schedule-based policy, a delivery is made on a regular basis (e.g., weekly). A hybrid policy combines the quantity-based and the schedule-based policies. This study aims to optimize ordering decisions in supply chain networks considering two order dispatch polices: quantity-based and schedule-based. For this purpose, this study introduces a Simulation-Optimization (SO) approach that combines Discrete-Event Simulation (DES) and steady-state Genetic Algorithm (ssGA). The SO is intended to find optimal set of network parameters including reordering policy parameters and dispatch process parameters, that minimize the total inventory cost and maintain a desirable customer service level.

## 2. Literature Review

Supply chain design and optimization problems are complex in nature. Garcia & You (2015) discussed four main challenges in addressing supply chain problems: large-scale modeling, large-scale optimization, uncertainty, and computational challenges. The modeling challenge appears when dealing with multi-stage integrated networks. In this case, models are required to cover detailed operations at different stages, and thus, optimization becomes very challenging when dealing with intractable mathematical models. The complexity of mathematical models grow rapidly as the level of detail grows. Alternatively, simulation can be used due to its capability of capturing intricate details. Simulation of supply chain operations can be used as a black box, by which different combinations of input parameters of the studied problem can be evaluated. The uncertainty is an important complexity feature of supply chain problems. DES can be used to mimic supply chain uncertainties of demand, delivery, and supply chain disruptions. Integrating simulation with optimization techniques allows for evaluating different combinations of decision variables to provide near-optimal solutions. This can be a practical substitute to other mathematical approaches such as stochastic programming (Macdonald et al., 2018; Oliveira et al., 2019). Details of using DES in optimization are available in (Swisher et al., 2000). The fourth challenge is the computational burden. Due to the size and uncertainty, solving mathematical supply chain models can be a challenge in terms of computational time. Thus, alternative optimization techniques such as metaheuristics can be used to obtain solutions in reasonable time. Abualigah et al. (2023) explores the application of metaheuristics in sustainable supply chain management. The study highlights the potential of these algorithms in enhancing both the sustainability and efficiency of supply chains. The study emphasizes the importance of selecting appropriate metaheuristic algorithms, considering problem complexity and data quality. Jalali & Nieuwenhuyse (2015) provide a review of SO application on supply chain problems. The authors reported that simulation combined with metaheuristic methods has been actively used in solving supply chain problems. They further reported that Genetic Algorithm (GA) is the most used metaheuristic for SO. In addition to their simplicity and effectiveness, evolutionary algorithms and GAs in particular are good for SO because of their capability of handling simulation noise (Jalali & Nieuwenhuyse, 2015). Example studies that involved GA-based SO are in (Gholizadeh & Fazlollahtabar, 2020; Ilgin & Tunali, 2007; Tsai & Fu, 2014; Vishnu et al., 2021). From the previous discussion, it can be concluded that SO represents a practical approach for tackling supply chain problems. With aid of simulation, SO is able capture high level of detail, account for uncertainty, and produce near optimal solutions in reasonable computational time. Simulation-optimization still interests researchers in the areas of constrained order fulfillment as seen in recent studies such as (Nanda & Patnaik, 2023).

The literature contains limited studies on optimized order dispatch management. Further, the available studies on order dispatch considered limited details of the process. These studies mainly focus on analyzing three order dispatch policies that were first studied by Çetinkaya et al. (2006). The policies can be quantity-based, schedule-based, or hybrid that combines the quantity-based and time-based policies. Studies of order dispatch policies include (Alnahhal et al., 2021; Bushuev, 2018; Howard & Marklund, 2011; Kang et al., 2017; Ongcunaruk et al., 2021; Rossetti & Liu, 2009; Wang & Takakuwa, 2006; Xiang & Rossetti, 2014). For instance, Jaruphongsa et al. (2004) studied optimal reordering and dispatch policies for a simple supply chain that includes single warehouse and single retailer considering delivery time windows. Çetinkaya et al. (2006) analyzed and compared the three policies based on total inventory cost. Xiang & Rossetti (2014) analyzed the effect of the policies on the lead time of demand orders. They also studied the effect on lead time of orders when a priority rule is applied on the dispatched orders. Kang et al. (2017) developed mathematical models for quantity-based and time-based policies and proposed algorithms using optimal properties of the models to compute the optimal parameters for ordering and delivery. Bushuev (2018) focused on enhancing supply chain delivery performance under delivery windows using a cost-based analytical model. The paper investigates various strategies aimed at optimizing delivery efficiency in two-stage supply chains, providing insights into cost-effective and performance-driven supply chain management. Al-Hawari et al. (2019) used DES to test the effect of different backlog and shipment rules on the total cost of multi-echelon supply chains. Alnahhal et al. (2021) used DES to investigate the effect of time-based temporal consolidation on the response time of outbound logistics under a soft delivery deadline. Ongcunaruk et al. (2021) developed a decision support tool that utilizes genetic algorithms to optimize delivery problems under mixed time window constraints.

This study extends the effort of optimizing ordering and dispatch processes to improve cost efficiency of supply chains and meet customer requirements. The approach in this study utilizes simulation to model a single warehouse multi-retailer network that includes order dispatch process, and delivery window constraints. In addition to the delivery window constraint, a delivery capacity constraint is imposed, where the warehouse can deliver at most one truckload per day. This study proposes a GA-based SO approach that works with combined total cost and service level objectives. The study further considers the details of the dispatch process through analyzing the tradeoff between a single dispatch queue and a dispatch queue per each order type (orders are classified by destination, e.g., retailer). Lastly, the study analyzes the impact of adopting a delivery priority rule for accumulated orders on total cost and service level of supply chains.

## 3. Problem Description

A two-echelon single-product supply chain network is considered in this study. The network includes a warehouse that fulfills the demand of a set of retailers. The warehouse refills its stock from an external supplier that's assumed to have infinite capacity. The warehouse and the retailers apply a continuous-review $(r, Q)$ ordering policy for inventory replenishment, where $r$ denotes the reorder point, and $Q$ denotes the reorder quantity. Under this policy, the inventory position (inventory on hand + inventory on order – backorders) of an item is continuously monitored. Whenever the inventory position falls to or under the reorder point, an order of fixed size $Q$ is placed (Federgruen & Zheng, 1992). When a retailer order arrives at the warehouse, it is satisfied immediately if the inventory on-hand is greater than or equal to the order size, otherwise, the whole order is placed in a backorder queue. When the warehouse replenishes its inventory, high priority is given to backordered retailer demands. This study focuses on the replenishment and delivery operations at the warehouse. The retailers here are viewed as a set of $m$ customers, where each retailer orders a fixed quantity $q_i$, $i = 1, 2, ..., m$. An important assumption here is that order split is not allowed in both fulfilling and delivering retailer orders. Thus, a certain retailer order of size $q_i$ is satisfied and delivered as a whole. The interarrival time between retailer orders is assumed to follow an exponential distribution with a rate $\lambda_i$.

After fulfilling retailer orders at the warehouse, the orders are accumulated into shipments via the dispatch process. Under a quantity-based policy, orders are accumulated until reaching a prespecified threshold that is less than or equal to a full truckload, assuming that the warehouse can only deliver one truckload a day. Whereas under a schedule-based policy, orders are accumulated for a prespecified number of days, after which, one truckload of orders is delivered. Figure 2 shows an illustration of the described supply chain network model.

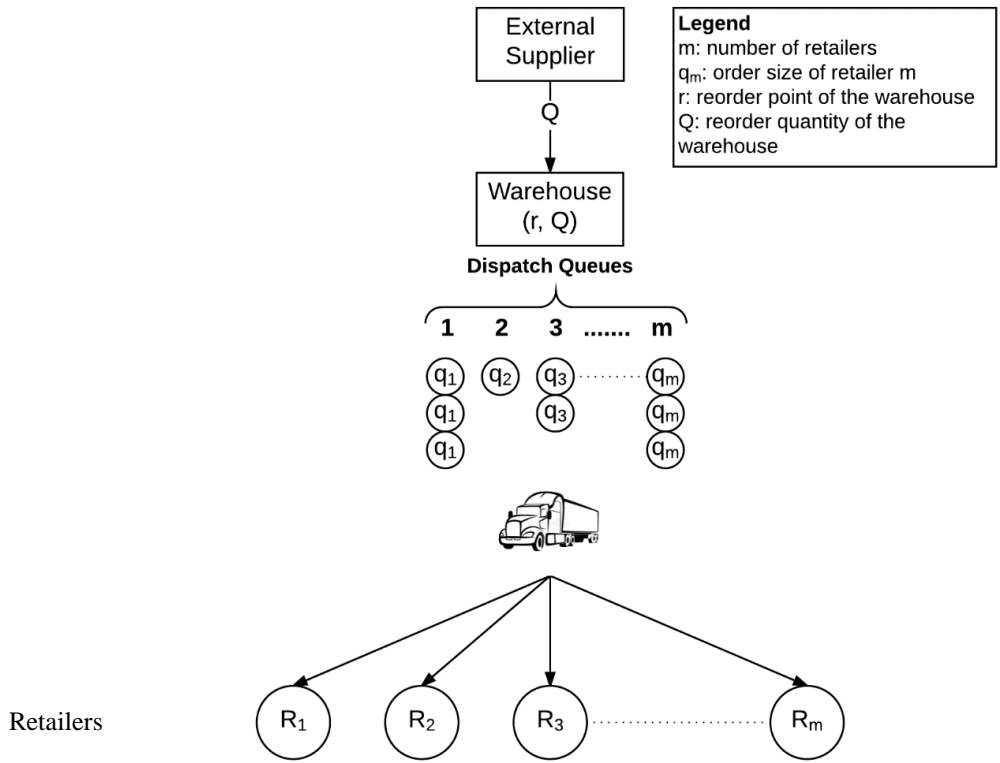

**Figure 2:** Supply chain network model with order dispatch process.

The warehouse is supposed to meet a certain delivery constraint, namely, a delivery time window $DW_i$ for each order $q_i$. If the warehouse violates the delivery window agreement either by making an early or a late delivery, a penalty cost is incurred. Early deliveries are also penalized as they may result with extra holding cost at the customer side, or handling issues when the customer is not ready to receive a delivery. Under the described delivery constraint, the warehouse should optimize its reordering and order dispatch policies in order to reduce the total cost and maintain a high customer service level.

## 4. Simulation-Optimization

4.1 Simulation Modeling

A DES model for the problem described in section 3 was implemented in Python 2.7 (Van Rossum & Drake Jr, 1995), using an event-scheduling approach. Detailed description of event-scheduling approach is provided in (Fishman, 2013). The model calculates two main performance measures: total cost and service level. The total cost includes holding, ordering, delivery, and penalty costs. The ordering cost is charged as a fixed amount per order and the delivery cost is also a fixed amount per delivery. Holding cost is proportional to the number of items at hand and the period of time they are held in stock. Holding cost rate ($h$) is a fixed amount charged per item per day spent in stock. Similarly, penalty cost is proportional to the number of items being delayed and the time of delay. Penalty cost rate ($p$) is a fixed amount charged per item per day of delay. Penalty costs include the backordering cost and the cost incurred by violating the delivery time window of orders.

The service level indicator considered in this study is the fill rate, which is the proportion of total retailer orders received by the warehouse that were satisfied from the inventory on-hand. The fill rate can be also viewed as the complement of stockout rate, which is the proportion of time an item was out of stock (Beamon, 1999).

Six versions of the simulation model were created to represent each of the scenarios addressed in this study. The scenarios represent different designs of the order dispatch process as shown in figure 3. The scenarios were generated by combining the three popular dispatch rules that were described in section 2 (quantity-based, schedule-based, and hybrid) with two additional dispatch management options. These two options are: (1) applying a priority rule of order dispatch, and (2) creating a single vs. multiple dispatch queues. The six scenarios are shown in shaded oval shapes in the figure. For the scenarios in which the dispatch is carried out using a single queue, this study investigates the effect of applying a priority rule of order dispatch, to replace the implicit First-In-First-Out (FIFO) priority rule. Under the Small-Orders-First (SOF) rule, smaller retailer orders are given higher delivery priority, regardless of how long the orders have been waiting in the dispatch queue.

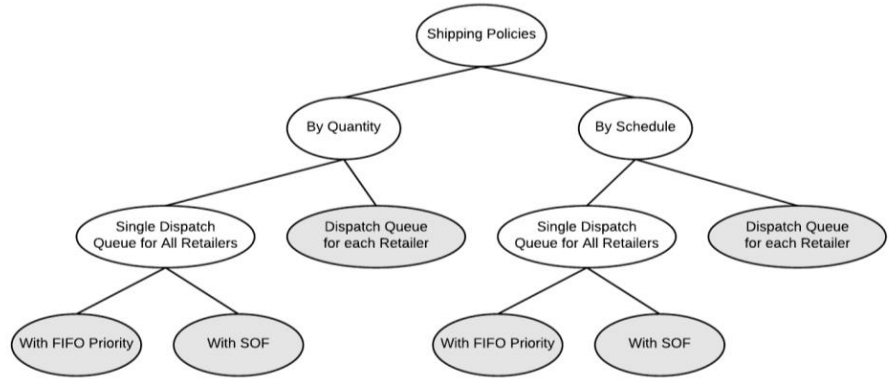

**Figure 3:** Different simulated scenarios of order dispatch process.

To verify the simulation models, different techniques were used such as plotting inventory variables, manual calculation of costs, real-time printing of dispatch queues, etc. For illustration, figure 4 shows a plot of the inventory position and the inventory on-hand at the warehouse in a 30-day pilot run of a 3-retailer, single-warehouse model.

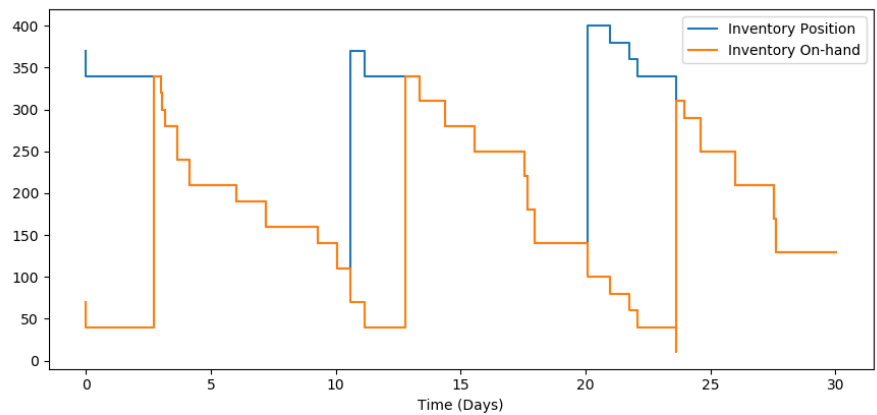

**Figure 4:** Inventory variables at the warehouse throughout a 30-day simulation run with r = 50 and Q = 350.

Run length was set to 100 days in all experiments. Each simulation run starts with a non-zero stock at the warehouse, which is initially set to a value equal to the reorder point. The number of simulation replicates ($n$) is dynamically determined during ssGA run using Algorithm 1. The purpose of Algorithm 1 is to determine $n$ such that a minimum desired precision of performance metrics is achieved. Precision in Algorithm 1 is defined as the ratio between the width of a 95% *t*-confidence interval ($W$) for a certain performance measure and the average ($\mu$) of the measure across replicates. As shown in steps 2.a through 2.c of the algorithm, simulations will be carried out until the ratio $W/\mu$ is less than or equal to a threshold value ($\delta$). The threshold in this study was set to 0.05 in all experiments. The minimum value for $n$ was set to 2 as shown in step 1 of the algorithm. Assuming that $n$ is determined based on the precision of total cost (*TC*), algorithm 1 can be used to stop the simulation after completing $n$ replicates.

---
**Algorithm 1:** Dynamic setting of the number of simulation replicates.

---
**Input:** $\delta$
**Output:** $\mu$, $W$
**Begin**
1. Run the simulation two times and record *TC* for each run
2. **Repeat:**
    2.a. Run the simulation one more time and record *TC*
    2.b. Calculate the average ($\mu$) divided by the width of 95% *t*-confidence interval (*W*) for all the recorded *TC* values
    2.c. **If** $W/\mu \leq$ d; go to 3
         **Else;** go to 2.a
3. **Return** $\mu$, $W$

**End**

---

Running the simulation with automatic determination of the number of replicates helps achieve desirable precision in each experiment, maintain consistent precision across experiments, and reduce the total number of runs needed.

4.2 Steady-State Genetic Algorithm

ssGA with integer-coded solutions is used in this study. In the implemented ssGA, one offspring is produced in each generation. The offspring solutions compete with the solutions of the parent population and the solutions with the top fitness values are selected for the next generation. The general procedure of the proposed ssGA is shown in Algorithm 2. Similar to the simulation models, ssGA was also implemented in Python 2.7 (Van Rossum & Drake Jr, 1995). The same procedure in section 4.1 (Algorithm 1) was used to determine the number of ssGA replicates.

---
**Algorithm 2:** ssGA procedure.

---
**Input:** Simulation model, ranges for decision variables, number of generations (*G*), mutation probability (*Pm*)
**Output:** Best solution found
**Start**
    *g = 0*
    Initialize: Randomly generate initial population $P^g$

---

```
            Evaluate fitness of P^g
            Best = best(P^g)
        While g < G, Do:
                Select 2 parents p_1 and p_2 from P^g
                Crossover p_1 and p_2 to produce offspring O^g
                Mutate O^g with probability Pm
                Evaluate O^g
                Replace O^g with the individual that has worst fitness in P^g
                If best(P^g) is better than Best
                        Best = best(P^g)
                g = g+1
        End While
        Return Best
End
```

As shown in Algorithm 2, ssGA procedure starts with initializing the search values and randomly generating $N$ solutions for an initial population, where $N$ is the population size. Fitness of the initial solutions is then evaluated and the initial global best (*Global_Best*) solution is identified. Afterwards, the ssGA search begins by running a prespecified $G$ number of generations per each algorithm run. In each generation, two parent solutions are randomly selected from the parent population for breading. Two breeding operators are applied. First, the selected parents are crossed over to generate two offspring solutions. Second, the two offspring solutions are mutated. The next step is to join the offspring and the parent population and then select the best $N$ solutions from the joined populations to advance to the next generation. In each generation, the surviving solutions are compared to the *Global_Best* solution to see if a better global solution was found. The following is a description of different components in the implemented ssGA:

- Decision variables and encoding

Decision variables include the ordering policy parameters ($r, Q$) in addition to the dispatch policy parameters at the warehouse. For the quantity-based policy, the dispatch parameter is a quantity threshold $M$ on the aggregate size of accumulated orders. The schedule-based policy has a dispatch parameter $S$, which is the number of days between two consecutive shipments. For multiple dispatch queues (a queue for each retailer), the number of dispatch policy variables is equal to the number of dispatch queues or the number of retailers. Thus, there will be $M_i$ and $S_i$ for $i = 1,...,m$, where $m$ is the number of retailers.

Solutions in the proposed ssGA are represented as chromosomes, where each gene carries a positive integer value to represent one decision variable. *r*, *Q*, and $S_i$ are allowed to take any integer value within their specified ranges. Because of the assumption that order split is not allowed, the quantity threshold $M_i$ is allowed to take integer values that are multiples of $q_i$, where $q_i$ is the order quantity of retailer *i*. Therefore, $q_i \leq M_i \leq truck\ capacity$.

- Fitness evaluation

The fitness function is used to evaluate and compare solutions throughout the ssGA search. ssGA runs the simulation model using the parameters in the solution (chromosome) to be evaluated and obtain two measures: average total cost and average fill rate at the warehouse. The total cost includes holding, ordering, delivery, and penalty costs. Fill rate is defined as the percentage of retailer orders satisfied from the on-hand inventory at the warehouse. The proposed fitness function combines the two objectives resulting in promoting solutions that minimize the total cost and maintain increase fill rate at the warehouse. The fitness function combines total cost and fill rate as follows:

$$F(p) = TC/FR \tag{1}$$

Where *F* is the fitness value of a solution *p*, *TC* is the average warehouse total cost, and *FR* is the is the average warehouse fill rate. By minimizing *TC* and maximizing *FR*, *F(p)* is minimized. The fill rate in equation (1) were allowed to take values greater than or equal to 0.01 to avoid division by zero.

- Selection

In each ssGA generation, two parent solutions are randomly selected for breeding. Tournament selection with a tournament size *k* = 3 is used for selecting each parent solution. In each tournament, the selection is performed as follows:

- Randomly sample three solutions from the population without replacement
- Evaluate the solutions fitness using equation (1)
- Rank the solutions according to their fitness in ascending order
- Select the solution in the top of the list (lowest fitness value)

- Crossover

Crossover is the first breeding operator applied on selected parent solutions and is performed in each generation (crossover rate = 1). Linear crossover is used with the variables *r, Q,* and *S$_i$*. Linear crossover calculates values for an offspring solution as linear combinations of the values in the two selected parent solutions. For illustration, if linear crossover is applied on the reorder point (*r*) values of two parent solutions, the *r* value of the offspring is calculated as follows:

$$r_o = r_1 * \alpha + r_2 * (1 - \alpha) \qquad (2)$$

Where *r$_o$* is the reorder point of the offspring, *r$_1$* and *r$_2$* are the reorder points of the parent solutions, and *α* is a uniformly distributed random number in the range [0, 1]. The values are rounded to the nearest integer after applying linear crossover to maintain integrality of the solution. Further, the result of the crossed over values is checked to make sure it falls within a prespecified range. If the values fall out of the range, they are brought back to the nearest extreme value.

Uniform crossover is applied on the solutions that have quantity-based dispatch parameters, *M$_i$*. It helps to maintain *M$_i$* values as multiples of *q$_i$*, due to the "no order split assumption" described earlier. The following illustrates a uniform crossover operation performed on *M$_i$* values:

- For each M$_i$ value in the offspring chromosome, Do
    - Generate a random number α = Uniform(0, 1)
        - If α > 0.5, set M$_i$ = M$_{i1}$
        - Else, set M$_i$ = M$_{i2}$

Where *M$_{i1}$* and *M$_{i2}$* are the delivery size thresholds in parent solutions 1 and 2, respectively.

- Mutation

Mutation is the second breeding operator applied on the offspring produced from crossing over two parent solutions. Mutation is applied on each gene in the offspring chromosome with a probability *Pm*. Gaussian mutation is applied on *r* and *Q* values by adding a value drawn from a normal distribution with $\mu = 0$ and $\sigma = 10$. The added value can be positive or negative and will either increment or decrement *r* and *Q* values. The mutated values are then rounded to the nearest integer and made sure they fall within their prespecified ranges. The values of schedule variables $S_i$ are mutated by randomly choosing between adding or subtracting one day. Similarly, $M_i$ variables are mutated by randomly choosing between incrementing or decrementing their values by a quantity equal to one order size $q_i$. In summary, the mutation operator is applied as follows:

- For each value in the offspring chromosome, Do
    - Generate a random number $\alpha$ = Uniform(0, 1)
        - If $\alpha$ < Pm, mutate the value (select between Gaussian or Uniform operator)
        - Else, move to the next value in the chromosome

## 5. Numerical Study

The SO approach described in section 4 was applied on one instance of the supply chain described in section 3. Data of the supply chain instance and parameters of the ssGA are described in sections 5.1 and 5.2, respectively. Results were obtained for the six scenarios described in figure 3. The results of these experiments are analyzed and discussed in section 5.3.

### 5.1 Simulation Model Data

The two-echelon network instance simulated in this study consist of one central warehouse and three retailers. The assumptions of ordering rates and ordering quantities for the network are as follows. The time between consecutive retailer orders that are sent to the warehouse is assumed to follow an exponential distribution with a mean arrival rate $\lambda$. Arrival rate of retailer orders from all the retailers combined is $\lambda = 1$ order/day. Hence, the three retailers have the same ordering rate $\lambda_1 = \lambda_2 = \lambda_3 = 1/3$ order/day. This implies that the mean interarrival time between two consecutive orders from the same retailer is 3 days. The reorder quantities ($q_i$) for each of the three retailers are assumed to be $q_1 = 50$ items, $q_2 = 100$ items, and $q_3 = 150$ items. These different retailer order sizes were set to allow for analyzing the cost impact of a quantity-based delivery priority rule. The ordering parameters for the warehouse ($r, Q$) are going to be optimized using ssGA. They are randomly selected for initial solutions. Ranges from which these parameters are selected are provided in section 5.2.

The warehouse operating costs and delivery parameters are summarized in table 1. Delivery cost is charged per truck trip regardless of the quantity being delivered. Ordering cost is a fixed charge per order. Holding cost is incurred per each day an item is held in inventory. Penalty cost is incurred per item per each day of delivery window violation, weather the item was delivered early or late. The delivery time window for all retailer orders is set as [3, 6], which implies that any retailer order must be delivered no earlier than 3 days, and no later than 6 days from the time it was placed.

**Table 1:** Warehouse operating costs and delivery parameters.

| Parameter | Value |
|---|---|
| Delivery cost ($/truck) | 500 |
| Ordering cost ($/order) | 200 |
| Holding cost rate ($/item/day) | 5 |
| Penalty cost rate ($/item/day) | 5 |
| Truck capacity (items) | 500 |
| Lead time of warehouse orders from the external supplier (days) | Uniform(2, 4) |
| Delivery time from warehouse to a retailer (days) | Uniform(2, 4) |
| Order delivery window at the warehouse (days) | [3, 6] |

This study analyzes the impact of establishing a single dispatch queue that ships to all retailers vs. the impact of establishing one dispatch queue per each retailer. If the dispatch process has multiple queues, each truckload released from the warehouse will include orders for one specific retailer only. Order delivery to any retailer is assumed to take *Uniform*(2, 4) days as indicated in table 1. However, if a single-queue dispatch process is applied, the warehouse will deliver to multiple retailers in the same trip. For simplicity, the travel times between the warehouse and each retailer and between each pair of retailers are assumed to follow *Uniform*(1, 2) days.

5.2 ssGA Parameter Setting

Table 2 shows the ranges for the decision variable, which define the search space of ssGA. The step size is also listed for each variable. For quantity-based policy parameters $M_i$, the upper bound is always equal to the truck capacity, while the lower bound is equal to the size of one retailer order $q_i$ in the case of multi-queue dispatch process. In case of single-queue dispatch, the lower bound of $M$ is set to the smallest possible retailer order size which is $q_1 = 50$ items.

**Table 2:** Lower and upper bounds of decision variables.

| Variable | Lower Bound | Upper Bound | Step Size |
|---|---|---|---|

| | | | | |
|---|---|---|---|---|
| $r$ (warehouse) | | 50 | 300 | 1 |
| $Q$ (warehouse) | | 200 | 1000 | 1 |
| Multi-queue quantity-based dispatch | | | | |
| $M_1$ | | 50 | 500 | 50 |
| $M_2$ | | 100 | 500 | 100 |
| $M_3$ | | 150 | 500 | 150 |
| Multi-queue schedule-based dispatch | | | | |
| $S_1$ | | 1 | 6 | 1 |
| $S_2$ | | 1 | 6 | 1 |
| $S_3$ | | 1 | 6 | 1 |
| Single-queue quantity-based dispatch | | | | |
| M | | 50 | 500 | 50 |
| Single-queue schedule-based dispatch | | | | |
| S | | 1 | 6 | 1 |

Initial ssGA population is randomly generated using the ranges in table 2. Table 3 shows the ssGA search parameters. The number of generations is fixed in all runs. ssGA parameters were tuned to provide reasonable search performance in terms of convergence, computational time, and solution quality. A pilot experiment was performed on one of the scenarios to examine the performance of the algorithm. Figure 5 shows a plot of the difference between the highest and lowest fitness values in the population for 1000 generations and five random initial populations. The plot shows a consistent performance of ssGA across the five runs and shows that the algorithm had sufficient number of generations to converge.

**Table 3:** ssGA parameters.

| Parameter | Value |
|---|---|
| Population size | 100 |
| Number of generations | 1000 |
| Crossover probability | 1 |
| Mutation probability | 0.2 |

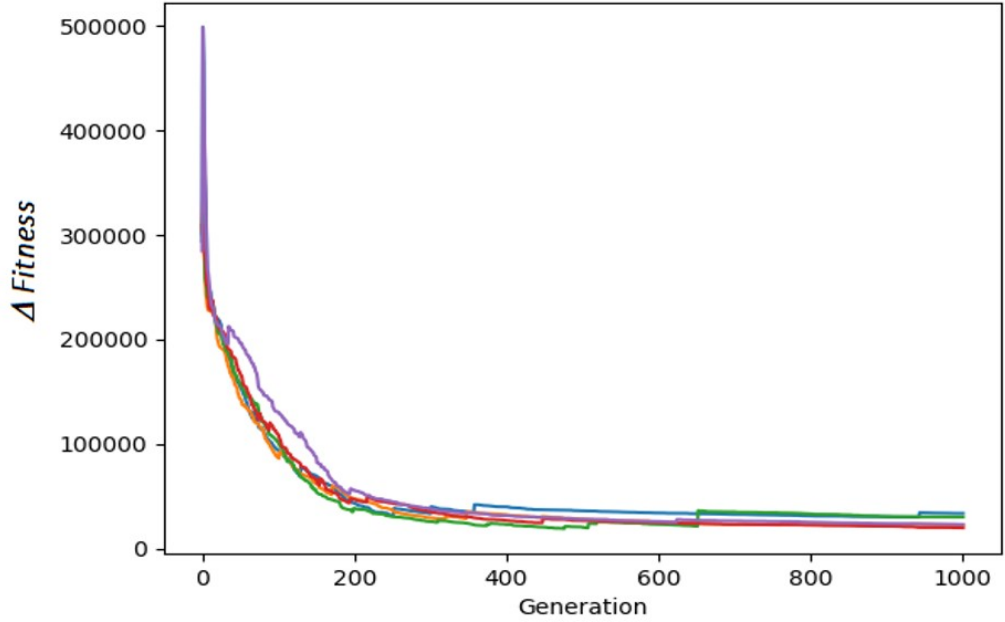

**Figure 5:** Difference between highest and lowest fitness values in the population for five ssGA runs.

5.3 Results and Discussion

In this section, the results obtained for the six simulation scenarios described in section 4.1 are analyzed and discussed. The scenarios are numbered from 1 to 6 along with their description in table 4. For each scenario, ssGA was run for *n* replicates, where *n* is different for each scenario. The average and 95% confidence interval of solution fitness, total cost, and fill rate were calculated for each scenario. The results are shown in figure 6.

**Table 4:** Description of Analyzed Scenarios.

| Scenario Number | Scenario Description |
|---|---|
| 1 | Quantity-based multi-queue dispatch |
| 2 | Quantity-based single-queue dispatch with FIFO priority |
| 3 | Quantity-based single-queue dispatch with SOF priority |
| 4 | Schedule-based multi-queue dispatch |
| 5 | Schedule-based single-queue dispatch with FIFO priority |
| 6 | Schedule-based single-queue dispatch with SOF priority |

Figure 6-c shows slight differences in the average fill rate across scenarios, which ranges between 76% and 80%. No significant difference in fill rate can be seen among different scenarios. Hence, it can be concluded that the different studied designs of order dispatch process do not have significant effects on the fill rate, when the reordering and dispatch process parameters are optimized. In other words, under different dispatch policies, a reasonable and consistent customer service level can be achieved by optimizing, reordering, and dispatch parameters. The same conclusion can be made when comparing the scenarios based on either solution fitness or total cost as shown in figure 6-a and 6-b.

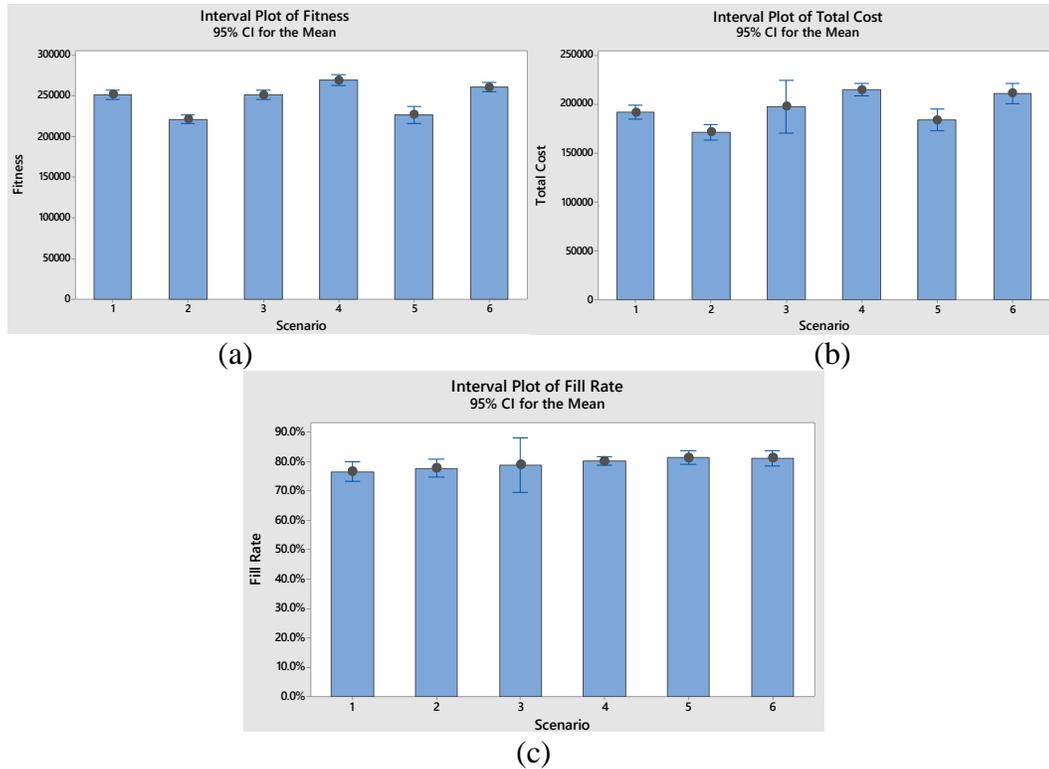

**Figure 6:** Average and 95% confidence interval of: (a) solution fitness, (b) total cost, and (c) fill rate across scenarios.

Scenario 2 resulted in the most cost saving as shown figures 6-a and 6-b. Scenario 2 involves a quantity-based single-queue dispatch policy without a priority rule. From figure 6, it can be further concluded that a single-queue quantity-based process with FIFO delivery priority outperforms the other studied scenarios, when the ordering and dispatch policies are optimized. However, with 95% confidence level, there was no significant difference in total cost and fitness value between the quantity-based and the schedule-based policies when using single-queue and FIFO priority. Further, when multiple dispatch queues are used, there was no significant difference in average fitness value between the quantity-based and the schedule-based policies. Based on the aforementioned observations, it can be noticed that comparing quantity-based and schedule-based policies under optimized parameters revealed no significant difference between the two policies in terms of cost saving.

Figure 6-a and 6-b also shows that a single-queue for dispatching orders results with a significant reduction in total cost, regardless of the dispatch method used (quantity or schedule). Moreover, regardless of the dispatch method used, replacing the FIFO delivery priority rule with SOF rule results in a significant increase of total cost at the warehouse. An interesting observation is that under both quantity-based and schedule-based policies, using multiple dispatch queues has the same effect as applying an SOF delivery priority rule on a single dispatch queue. By dispatching orders of the three retailers using separate queues, the warehouse is expected to deliver more frequently compared to the single-queue case. This helps in avoiding penalty costs incurred from delayed deliveries, however, delivery cost will increase. For a single-queue case, when SOF rule is applied, the orders of retailer 1 will be delivered faster than those of retailers 2 and 3. According to the delivery window agreement, deliveries must not be made sooner than 3 days from the time of delivery. Thus, some of retailer 1 orders might be penalized. Since the order size of retailer 3 is the largest, a delayed delivery to retailer 3 can incur a substantial penalty cost. These penalty costs appear to be equivalent to the extra delivery cost incurred when multiple dispatch queues are used.

Since the average fill rate is consistent across scenarios, the best solution reached in each scenario was selected based on the average total cost. The best solutions for the six scenarios are shown in table 5. The lowest achieved total cost is $157,486.5 for the single-queue quantity-based policy with FIFO rule. This cost was achieved by setting the reorder point to 303 items, reorder quantity to 261 items, and the threshold on dispatch quantity to 300 items.

**Table 5:** Best solution found for each scenario.

| Scenario | Average Total Cost | r | Q | Dispatch Policy Parameters |
|---|---|---|---|---|
| 1 | 184,058.5 | 285 | 299 | $M_1 = 100$, $M_2 = 200$, $M_3 = 150$ |
| 2 | 157,486.5 | 303 | 261 | $M = 300$ |
| 3 | 191,430.0 | 255 | 350 | $M = 300$ |
| 4 | 192,713.1 | 277 | 316 | $S_1 = 3$, $S_2 = 3$, $S_3 = 3$ |
| 5 | 165,103.5 | 286 | 275 | $S = 2$ |
| 6 | 200,840.6 | 285 | 335 | $S = 1$ |

Although the maximum allowed delivery size is 500 (truck capacity), it can be noticed that none of the optimized thresholds (*M* values) reached close to 500. Due to the delivery window constraint, the warehouse makes deliveries more frequently with sizes less than 500 to avoid high penalty cost on untimely deliveries. Increasing the frequency of delivery will increase the delivery cost. However, the optimized *M* values help creating a balance between potential penalty and delivery costs, in order to minimize the total cost at the warehouse. For scenario 1, the results suggest that the warehouse should make a delivery to retailer 1 when 2 orders are accumulated ($M_1 = 100 = 2*q_1$, $q_1 = 50$), and the same for retailer 2. However, the warehouse must deliver retailer the three orders individually.

The multi-queue policy results in more frequent deliveries as pointed out earlier, and this increases the total delivery cost. In the case of schedule-based policy, the warehouse must make a delivery to each retailer every three days in the multi-queue case. However, the total cost decreases in the single-queue case, where the warehouse releases one truckload every two days. By comparing scenarios 2 and 3, the *M* value did not change when the SOF priority rule was applied. However, the total cost increased due to the penalty costs resulted from delaying low priority large-size orders.

For scenarios 5 and 6, adding the SOF priority rule will increase the frequency by decreasing the time between two deliveries, *S,* from 2 to 1. Under the SOF rule, the warehouse is making deliveries on a daily basis in order to minimize order delays and reduce penalties.

By inspecting the best *r* and *Q* values in table 6, it can be noticed that the values are quite close to each other across the scenarios. Figure 7 shows a plot of the average and 95% confidence interval of *r* and *Q* values obtained for each scenario. The plot shows that there is no significant difference between *r* values as the dispatch policy changes. The same conclusion can be made for Q values, except for the slight difference that can be seen between scenario 2 and 4.

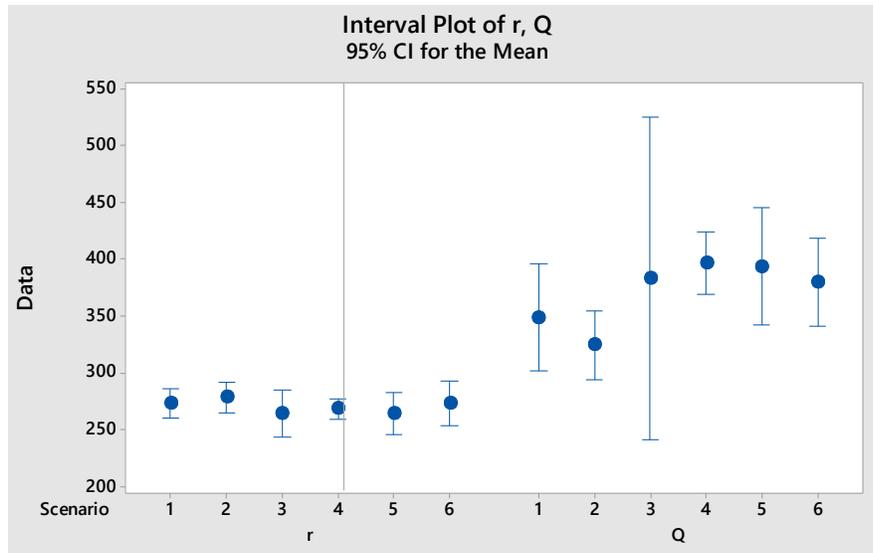

**Figure 7:** Average and 95% confidence interval of r and Q values obtained for each scenario.

*r* and *Q* values impact holding, ordering, and backordering costs. However, there is no evidence in the studied cases that *r* and *Q* values have direct impact on the delivery cost or the penalty costs associated with order wait time in the dispatch queues. Optimal *r* and *Q* values are mainly influenced by the demand characteristics of the retailers such as order quantities and interarrival time of orders. If the details of ordering policies at the retailers are considered, the dispatch process may start to influence the tuning of *r* and *Q* at the warehouse. When the lead time of retailer orders from the warehouse increases due to transportation constraints and dispatch policies, the retailers may increase their reorder quantities and ordering frequency to maintain their service levels. Thus, the studied supply chain model can be expanded to include reordering policies at the retailer level. This would help in exploring the interaction between ordering policies at the retailers and the warehouse under different order dispatch policies.

## 6. Managerial Insights

The findings from our simulation scenarios provide key insights for supply chain managers regarding the optimization of order dispatch policies. Scenario 2, featuring a quantity-based single-queue dispatch policy with FIFO priority, emerged as the most cost-effective approach as per the numerical analysis discussed in section 5. This scenario can lead to potential cost savings without compromising the service level. However, an essential observation is that the type of dispatch policy (quantity-based vs. schedule-based) did not significantly impact the total cost or service level when optimized parameters were used. This insight challenges the traditional preference for one policy over the other, suggesting that a well-optimized dispatch system, regardless of its base policy, can yield comparable cost savings. Managers should note that a single-queue dispatch approach, irrespective of the method used, may reduces total costs, offering a practical solution for optimizing supply chain efficiency.

The results discussed in section 5 also highlight the strategic importance of choosing appropriate dispatch process parameters. For instance, using multiple dispatch queues or applying an SOF delivery priority rule leads to increased delivery frequency and consequently higher total costs. These insights can inform decisions about the number of dispatch queues and the application of priority rules. The study emphasizes the balance between potential penalty costs due to delayed deliveries and increased delivery costs. Further, the optimized reorder points and quantities ($r$ and $Q$ values) suggest that understanding the demand characteristics of retailers, like order quantities and interarrival times, is crucial. This knowledge can be leveraged to tune the dispatch process effectively, leading to optimized operational costs. The findings advocate for an expansion of the supply chain model to include reordering policies at the retailer level, offering a comprehensive perspective on the interaction between ordering policies at different supply chain stages and their impact on overall efficiency and cost.

## 7. Conclusion

This study aimed to identify optimal reordering and order dispatch policies in a two-echelon supply chain network constrained by delivery windows. The network comprises a group of retailers serviced by a single warehouse. The warehouse consolidates completed retailer orders into shipments of economical size based on an order dispatch policy, which may be either quantity-based or schedule-based. The research introduced a Genetic Algorithm-based Simulation-Optimization (GA-based SO) approach that targets combined total cost and service level objectives for network optimization. It analyzed various implementations of order dispatch policies, including single and multiple dispatch queues, and investigated the impact of applying a delivery priority rule to the accumulated orders.

The findings indicate that a single-queue, quantity-based dispatch policy with FIFO delivery priority leads to the most significant reduction in total cost at the warehouse. Nevertheless, under a single-queue and FIFO rule, there was no notable difference in total cost between the quantity-based and schedule-based policies. This suggests that deeper insights into dispatch policy performance can be gained by examining the detailed aspects of the dispatch process. For the quantity-based policy, the optimized threshold, denoted as M, successfully struck a balance between delivery costs and potential penalty costs, thereby minimizing the warehouse's total cost.

The analysis of the optimized reordering parameters ($r$, $Q$) revealed that both $r$ and $Q$ values remained fairly consistent across the various scenarios tested. Given that these values are largely influenced by the retailers' ordering behavior, no evidence was found of an interaction between reordering and dispatch policies. Future research could benefit from expanding the network model to encompass customer demand and reordering operations at the retailer level, to further explore possible interactions.

Ongcunaruk, W., Ongkunaruk, P., & Janssens, G. K. (2021). Genetic algorithm for a delivery problem with mixed time windows. *Computers & Industrial Engineering*, *159*, 107478. https://doi.org/10.1016/j.cie.2021.107478

Rossetti, M. D., & Liu, Y. (2009). Simulating SKU Proliferation in a Health Care Supply Chain. *Winter Simulation Conference*, 2365–2374. http://dl.acm.org/citation.cfm?id=1995456.1995779

Swisher, J. R., Hyden, P. D., Jacobson, S. H., & Schruben, L. W. (2000). A survey of simulation optimization techniques and procedures. *Simulation Conference, 2000. Proceedings. Winter*, *1*, 119–128.

Tsai, S. C., & Fu, S. Y. (2014). Genetic-algorithm-based simulation optimization considering a single stochastic constraint. *European Journal of Operational Research*, *236*(1), 113–125. https://doi.org/10.1016/j.ejor.2013.11.034

Van Rossum, G., & Drake Jr, F. L. (1995). *Python tutorial*. Centrum voor Wiskunde en Informatica Amsterdam, The Netherlands.

Vishnu, C. R., Das, S. P., Sridharan, R., Ram Kumar, P. N., & Narahari, N. S. (2021). Development of a reliable and flexible supply chain network design model: A genetic algorithm based approach. *International Journal of Production Research*, *59*(20), 6185–6209. https://doi.org/10.1080/00207543.2020.1808256

Wang, X., & Takakuwa, S. (2006). Module-Based Modeling of Production-Distribution Systems Considering Shipment Consolidation. *Simulation Conference, 2006. WSC 06. Proceedings of the Winter*, 1477–1484. https://doi.org/10.1109/WSC.2006.322916

Xiang, Y., & Rossetti, M. D. (2014). The effect of backlog queue and load-building processing in a multi-echelon inventory network. *Simulation Modelling Practice and Theory*, *43*, 54–66. https://doi.org/10.1016/j.simpat.2014.01.006
24